\documentclass[preprint2]{aastex} 

\def\ltsima{$\; \buildrel < \over \sim \;$}
\def\simlt{\lower.5ex\hbox{\ltsima}}
\def\gtsima{$\; \buildrel > \over \sim \;$}
\def\simgt{\lower.5ex\hbox{\gtsima}}

\slugcomment{Submitted to ApJ, 1999 December 23}

\begin{document}

\title{The Central X-Ray Point Source in Cassiopeia A}

\author{Deepto~Chakrabarty and Michael~J.~Pivovaroff}
\affil{\footnotesize Department of Physics and Center for Space Research, 
   Massachusetts Institute of Technology, Cambridge, MA 02139}
\email{deepto@space.mit.edu, mjp@space.mit.edu}
\and

\author{Lars~E.~Hernquist, Jeremy S. Heyl,\altaffilmark{1},
        and Ramesh Narayan}
\affil{\footnotesize Harvard-Smithsonian Center for Astrophysics, 
   60 Garden Street, Cambridge, MA 02138}
\email{lhernqui@kona.harvard.edu, jheyl@cfa.harvard.edu, 
rnarayan@cfa.harvard.edu}

\author{\ }
\affil{{\sc The Astrophysical Journal}, {\rm Vol. 546, in press (2001 January 10)}}
\affil{Submitted 1999 December 23; accepted 2000 August 29.}

\altaffiltext{1}{Also Theoretical Astrophysics, California Institute of
   Technology, Pasadena, CA 91125}

\begin{abstract}
The spectacular ``first light'' observation by the {\em Chandra X-Ray
Observatory} revealed an X-ray point source near the center of the 300
yr old Cas~A supernova remnant.  We present an analysis of the public
X-ray spectral and timing data.  No coherent pulsations were detected
in the {\em Chandra}/HRC data.  The 3$\sigma$ upper limit on the
pulsed fraction is $<$35\% for $P>20$ ms.  The {\em Chandra}/ACIS
spectrum of the point source may be fit with an ideal blackbody
($kT$=0.5 keV), or with blackbody models modified by the presence of a
neutron star atmosphere ($kT$=0.25--0.35 keV), but the temperature is
higher and the inferred emitting area lower than expected for a 300 yr
old neutron star according to standard cooling models.  The spectrum
may also be fit with a power law model (photon index
$\Gamma=2.8$--3.6).  Both the spectral properties and the timing
limits of the point source are inconsistent with a young Crab-like
pulsar, but are quite similar to the properties of the anomalous X-ray
pulsars.  The spectral parameters are also very similar to those of
the other radio-quiet X-ray point sources in the supernova remnants
Pup~A, RCW~103, and PKS~1209--52.  Current limits on an optical
counterpart for the Cas~A point source rule out models that invoke
fallback accretion onto a compact object if fallback disk properties
are similar to those in quiescent low-mass X-ray binaries.  However,
the optical limits are marginally consistent with plausible
alternative assumptions for a fallback disk.  In this case, accreting
neutron star models can explain the X-ray data, but an accreting black
hole model is not promising.
\end{abstract}

\keywords{accretion, accretion disks --- black hole physics --- stars:
neutron --- stars: peculiar --- supernova remnants --- supernovae: 
individual (Cas~A)}

\section{INTRODUCTION}

For over three decades, it has been well established that (some)
supernova explosions give rise to strongly magnetized ($B\sim 10^{12}$
G), rapidly rotating ($P\sim10$--30 ms) neutron stars (NSs), as in the
young radio pulsars found in the Crab Nebula and nearly a dozen other
supernova remnants (SNRs).  In some cases, a synchrotron nebula
(or ``plerion'') has been detected around the pulsar, powered by
non-thermal emission from the NS.  Emission (in some cases pulsed) has
been detected at other wavelengths (optical, X-ray, gamma-ray) arising
from thermal and non-thermal processes.  However, several clues have
recently emerged suggesting that this paradigm is incomplete (see
Kaspi 2000 and Gotthelf \& Vasisht 2000 for recent reviews).  First,
there are the six slowly-rotating ($P\sim6$ s) ``anomalous X-ray
pulsars'' (AXPs), which seem to be young isolated NSs and may have
extremely strong ($B\sim 10^{14}$--$10^{15}$ G) surface magnetic
fields (Mereghetti 2000).  Half of the AXPs are associated with SNRs.
Possibly related are the four known soft gamma-ray repeaters (SGRs),
which in quiescence share many properties with AXPs and may also be
associated with SNRs (Hurley 2000).

Also intriguing has been the identification of at least three radio-quiet
non-plerionic X-ray point sources near the centers of SNRs (Pup A,
RCW~103, and PKS 1209--52; see Brazier \& Johnston 1999 and references
therein).    These objects have X-ray 
spectra roughly consistent with young, cooling NSs, but show no
evidence for either X-ray pulsations or emission at other wavelengths
(in contrast to ``normal'' young NSs).   Finally, the ongoing failure
to detect clear evidence for a young NS in the remnant of SN 1987A in
the Large Magellanic Cloud (LMC) has renewed theoretical interest in
alternative models for the aftermath of a SN explosion, especially
with respect to fallback of ejected material onto a newborn NS.
Several groups have concluded that, under some circumstances, a
newborn NS might collapse into a black hole (BH) shortly after birth
(Brown \& Bethe 1994;  Woosley \& Timmes 1996; Zampieri et al. 1998;
Fryer, Colgate, \& Pinto 1999). 

Nearby SNRs without known stellar remnants are thus obvious targets
for further study.  After SN 1987A, the youngest known SNR in our
Galaxy or the satellite Magellanic Clouds is Cassiopeia A.  Its parent
supernova was evidently noticed (though misunderstood) by Flamsteed in
1680 (Ashworth 1980).  This $\tau_{\rm hist}=320$ yr historical age
for Cas~A is consistent with its optical expansion time scale (van den
Bergh \& Kamper 1983), though somewhat shorter than its X-ray
($\tau_{\rm x}\approx 500$ yr; Koralesky et al. 1998, Vink et
al. 1998) and radio ($\tau_{\rm radio}\sim$750--870 yr; Anderson \&
Rudnick 1995) expansion time scales.  The progenitor of this
oxygen-rich SNR was probably a very massive (zero-age main sequence
mass $M_{\rm ZAMS}>20 M_\odot$) late WN-type Wolf-Rayet star which
underwent prodigious mass loss via a stellar wind and eventually
exploded as a type II supernova (Fesen, Becker, \& Blair 1987).  The
inferred distance to Cas~A is $3.4^{+0.3}_{-0.1}$ kpc (Reed et
al. 1995).  The remnant subtends 4 arcmin in the sky, is one of the
brightest non-thermal radio sources, and has been extensively
studied in the radio, optical, and X-ray bands.

The {\em Chandra X-Ray Observatory} ``first light'' observation on
1999 August 20 revealed the presence of a compact X-ray source near
the geometric center of Cas~A (Tananbaum 1999).  The source morphology
is point-like, with no obvious evidence for extension or a surrounding
nebula (e.g., a plerion).  The discovery announcement notes that no
obvious counterparts were detected within a 5 arcsec radius of the
point source position on 20 cm radio maps or optical images.  Previous
X-ray missions lacked sufficient spatial resolution to separate the
point source from the diffuse SNR emisison.   However, armed with the
{\em Chandra} position, Aschenbach (1999) detected the point source in
archival 0.1--2.4 keV X-ray images taken with the {\em ROSAT}/HRI in
1995-1996, and Pavlov \& Zavlin (1999) recovered the point source in
archival 0.5--4 keV X-ray images taken with the {\em Einstein}/HRI in
1979 and 1981.  The {\em Einstein}, {\em ROSAT}, and {\em Chandra}
count rates were consistent with a constant X-ray source flux over all
the observations (Pavlov \& Zavlin 1999; Pavlov et al. 2000).  They
also noted that the observed spectrum appeared to be inconsistent with
pure blackbody radiation from the entire surface of a cooling NS.
Umeda et al. (2000) speculated on some possible scenarios for the
nature of the point source, based on these early results.

In this paper, we present a detailed analysis of the X-ray spectral and
timing features of the central point source in Cas~A, based on the
available public {\em Chandra} data.   In \S2, we give a detailed
description of the observation and our data analysis, including our
efforts to verify the instrumental calibration.  In \S3, we discuss
our results in the context of various models for the nature of the
point source.  We summarize our finding in \S4.  While completing our
manuscript, we learned of another paper presenting an independent analysis
of the spectral data by Pavlov et al. (2000).  They used a subset of
the data that we discuss in our paper, and their spectral results are 
consistent with ours within the uncertainties.  We include a brief
discussion of their preferred interpretation in \S3. 

\section{OBSERVATIONS AND ANALYSIS}

{\em Chandra} (formerly {\em AXAF}; Weisskopf et al. 2000)
was launched on 1999 July 23.  Its High Resolution
Mirror Assembly (HRMA), which consists of four pairs of nested
grazing-incidence Wolter type I mirrors with a 10~m focal length,
focuses X-rays in the 0.1--10 keV range.  The on-axis point-spread
function of the HRMA has a 50\% energy width of 0.25 arcsec at 0.3 keV
and 0.6 arcsec at 9.7 keV.  Numerous imaging observations of Cas~A
have been made by {\em Chandra} as part of the mission's Orbital
Activation and Checkout (OAC) calibration program.  All OAC data are
immediately in the public domain.  Most of the Cas~A observations were
made using the Advanced CCD Imaging Spectrometer (ACIS; Burke et
al. 1997), which records both the sky position (0.49 arcsec/pixel) and
the energy ($\Delta E \approx$ 50--200 eV) of each detected photon in
the 0.1--10 keV range, with a time resolution of 3.2 s.  A few of the
observations were made with the High Resolution Camera (HRC; Zombeck
et al. 1995; Murray et al. 1997), which precisely records the sky
position (0.13 arcsec/pix) and arrival time\footnote{A recently
identified wiring problem in the HRC onboard electronics has caused
the event timing accuracy to be degraded to $\sim$4 ms for the data
presented in this paper (S.~S. Murray 2000, private communication;
Seward 2000).  See \S2.3 for further discussion.} ($\Delta t$=16
$\mu$s) of each detected photon, but with modest ($E/\Delta E \sim 1$)
energy resolution.  No diffraction gratings were in place for any of
the ACIS or HRC observations.  A summary of the observations used in our
analysis is given in Table~1. 
\begin{deluxetable}{ccccccccc}
\setlength{\tabcolsep}{0.04in}
\tabletypesize{\small}
\tablewidth{0pt}
\tablecaption{Selected {\em Chandra} Observations of Cas~A}
\tablehead{
  &  & & \colhead{$\theta_{\rm off-axis}$} & \colhead{$R_{\rm aper}$} &
   \colhead{$T_{\rm elapsed}$} & 
   \colhead{$T_{\rm good}$} & 
   \colhead{$N_{\rm aper}$\tablenotemark{a}} & 
   \colhead{Rate\tablenotemark{b}}\\
\colhead{ObsID} & \colhead{Start time (UT)} & \colhead{Instrument}& 
   \colhead{(arcmin)}& \colhead{(arcsec)}& \colhead{(s)} & \colhead{(s)} & 
   \colhead{(ct)}&\colhead{(ct ks$^{-1}$)}}
\startdata
214&1999 Aug 20, 00:07&ACIS/S3&2.3& 3.3 & 6107 & 2803 & 420 &$115\pm7$\\
220&1999 Aug 22, 23:27& ACIS/S3 & 2.8 & 2.0 & 4279 & 1212 & 150 &$113\pm10$\\
221&1999 Aug 23, 00:52& ACIS/S3 & 3.9 & 4.5 & 2103 & 1060 & 199 &$128\pm13$\\
222&1999 Aug 23, 01:38& ACIS/S3 & 2.8 & 3.8 & 2080 & 1044 & 179 &$125\pm13$\\
 \\
172 & 1999 Sep 05, 18:45 & HRC-S & 1.3 & 3.0 & 9485 & 9485 & 411 & $43\pm2$\\
1409& 1999 Oct 23, 18:31 & HRC-I & 1.4 & 3.0 & 12770 &12770 & 479 &$38\pm2$\\
\enddata
\tablenotetext{a}{Counts in extracted aperture (including background).}
\tablenotetext{b}{Point source count rate (background subtracted).}
\end{deluxetable}

\subsection{ACIS Data Reduction}

As the analysis tools and calibration for {\em Chandra} are still
under development at this early stage of the mission, we will describe
our data reduction, analysis, and verification steps in detail.
In using the ACIS data to derive a spectrum for the point source in
Cas~A, we have restricted our analysis to the four exposures obtained
during 1999 August 20--23 with Cas~A placed on the back-illuminated
(BI) ACIS S3 chip, for three reasons.  First, preliminary analysis
showed that the point source spectrum was relatively soft, and the two
BI chips (S1 and S3) have superior low energy response compared to the
front-illuminated (FI) chips.  Second, we wished to optimize the {\em
Chandra} point spread function by minimizing the off-axis angle of the
point source, and the {\em Chandra} aim points for ACIS lie on either
the S3 or I3 chips.  Inspection of the Cas~A images obtained on the
other ACIS chips shows that the point source is spread out over many
more pixels, making it more difficult to separate from the background; the
effective area of the mirror and detector combination at these angles
is also reduced.  Finally, the BI chips have not suffered the
radiation damage that degraded the energy resolution of the FI chips
soon after launch, and so are better calibrated at this early stage in
the mission.   A summary of the observations we used is given in Table~1.  

We obtained the fully processed (level 2) ACIS event data for these
observations from the {\em Chandra} Data Archive.  The data were
acquired in Timed-Exposure/Faint mode, with a frame read out every
3.2~s.  A substantial number of frames was lost to telemetry
saturation due to the high total count rate from Cas~A, resulting in
an overall observation duty cycle of 42\%.  We filtered the events
from the surviving frames, accepting only those which fell within the
``standard'' event grade set (grades 0+2+3+4+6) in order to maximize
the ratio of X-ray to non--X-ray events, and additionally discarding
events with very large pulse heights as due to cosmic rays.  For all
the observations, the ACIS focal plane temperature was $-100^\circ$~C.

We did not attempt to improve upon the spacecraft aspect solutions
from the standard processing, but instead proceeded from the
assumption that the central source in Cas~A is indeed point-like in
morphology, as determined by Tananbaum (1999).  For each observation,
we extracted all events located within a given radius of the centroid
position of the point source.  Ideally, we would use the calibrated
angular response of the HRMA to choose an extraction radius which
encircled some fixed fraction (e.g., 95\%) of the flux from a point
source for a given off-axis angle.  However, both the focus position
of the detector and the quality of the spacecraft aspect solution were
not necessarily optimal in these early observations.  Instead, we
measured the radial distribution of source photons for each
observation from the data and estimated a 95\% extraction radius.
Table~1 lists the aperture sizes.

The background in the point source region is dominated by the diffuse
emission of the supernova remnant and has strong contributions from
both continuum and line emission (e.g., Holt et al. 1994).  The point
source is located within a relatively low surface brightness region of
the remnant.  Still, the background must be estimated with care, as
there are significant compositional gradients in this part of the
remnant (and perhaps variations in the underlying plasma conditions),
resulting in varying line strengths with position (see also Hughes et
al. 2000).  After investigating the line strengths in a number of
nearby regions, we selected an off-centered $48\times15$ arcsec$^2$
rectangular region around the point source (with the point source
itself and a small surrounding buffer region excluded) to estimate the
background surface brightness near the point source.  The point source
and the background region are shown in Figure 1.
\begin{figure}
\figurenum{1}
\plotone{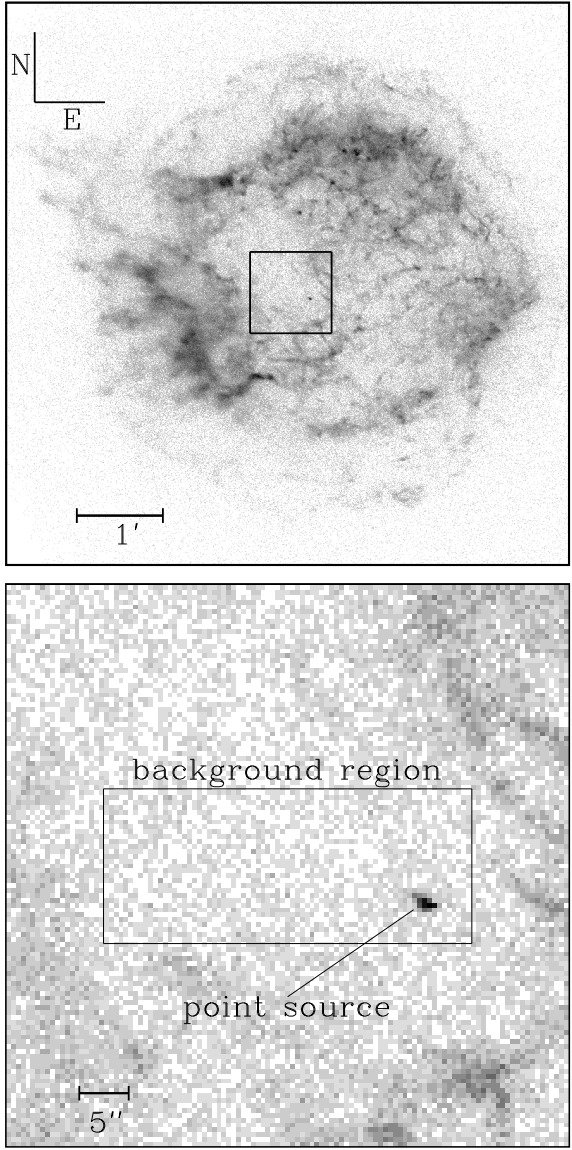}
\caption{A broad-band 0.3--10~keV intensity map of
Cassiopeia~A, as imaged by a 2.8~ks exposure with ACIS S3 (ObsID~214).
The upper panel shows the intricate structure present, including the
central point source.  The lower panel is an expanded view of the
boxed region in the upper panel.  Both the point source and the
rectangular region used for the background region (see $\S$2.2) are
indicated.}
\end{figure}

The ACIS instrumental response may be parameterized by a photon
energy redistribution matrix (hereafter RMF), which maps incident
photon energy to detected pulse height amplitudes (PHAs).   Every ACIS
chip ($1024\times 1026$) is divided into four parallel ($256\times 1026$)
quadrants, each read out separately by four individual amplifiers.
Each quadrant requires a unique RMF, as each readout amplifier has a
different gain.  Additionally, the parallel and serial charge transfer
inefficiency in the BI chips leads to differences in energy resolution
and detection efficiency versus position within a quadrant as well.
Consequently, the ACIS instrument team has supplied multiple
position-dependent RMFs for each quadrant.  We used the 1999 October
28 version of RMFs developed for observations taken at a focal plane
temperature of $-100^\circ$ C.

Fortuitously, the Cas~A point source (and nearly all of the selected
background region) lies entirely in quadrant 2 for all four of the S3
observations.  Furthermore, the four observations were located within
250 pixels of each other within the chip quadrant, minimizing the
differences in CCD response.  Thus, the four RMFs appropriate for each
observation may be combined without any serious loss of accuracy. 
This allows us to sum the four individual S3 count spectra into a
single ``grand total'' S3 count spectrum with improved statistics. 
In addition to the RMFs, an ancillary response function (hereafter
ARF) is needed to characterize the effective area of the HRMA and the
quantum efficiency of the ACIS detector as a function of incident
photon energy and off-axis angle.  We computed the appropriate ARF for
each observation using software created by members of the ACIS and
High Energy Transmission Grating instrument teams at MIT.  The overall
instrumental response is given by the product of the RMF and the ARF.
We computed this product for each observation, and then combined these
products by an exposure-weighted average for use with our summed count
spectrum.   

\subsection{X-Ray Spectral Fitting}

The summed, background-subtracted ACIS count spectrum was analyzed
using the XSPEC v10.00 software package (Arnaud 1996).  We rebinned
the data at 0.15 keV resolution (32 ADU/channel) and restricted our
analysis to 21 bins in the 0.7--4.0 keV range, resulting in at least
10 count bin$^{-1}$ in all but 3 bins at high energy (with a minimum
of 5 count bin$^{-1}$ and a maximum of 71 count bin$^{-1}$), and a
total of 630 counts.  The rebinned count spectrum is shown in Figure~2.
\begin{figure}
\figurenum{2}
\plotone{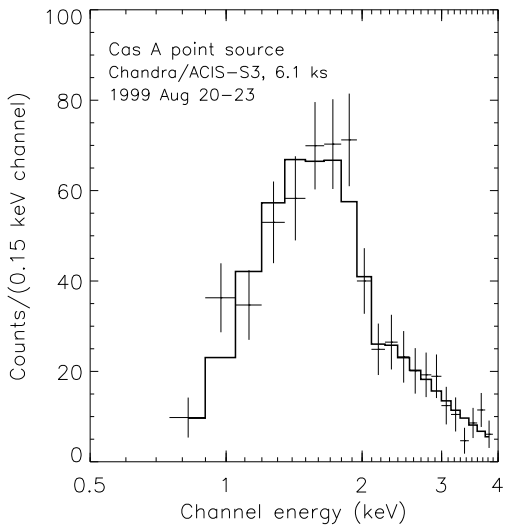}
\caption{The background-subtracted {\em Chandra}/ACIS-S
count spectrum of the Cas~A point source, summed from the four
individual S3 exposure on 1999 Aug 20--23 for a total exposure of
6119~s.  The dark line shows the spectrum predicted by an absorbed
blackbody model with $kT^\infty=0.53$ keV and $N_H=0.84\times 10^{22}$
cm$^{-2}$.  Other models (power law, NS atmosphere) may be similarly fit;
the present data do not discriminate strongly between these models.}
\end{figure}
In light of the relatively low count rate, our model fitting
weighted each bin by $1+(N_i + 0.75)^{1/2}$ rather than $N_i^{1/2}$
(where $N_i$ is the number of counts in bin $i$).  This is a superior
weighting scheme when $N_i$ is small and asymptotes to the usual
weighting when $N_i$ is large (Gehrels 1986).

We fit several different spectral models to these data: a simple power
law, thermal bremsstrahlung, an ideal blackbody (BB), and two modified
BB models\footnote{For the BB and modified BB models, the quoted
spectral parameters are those that would be measured by an observer at
infinity, assuming the gravitational redshift correction factor for a 
10 km NS.}.  The modified BB models assume that the emitting object is a NS
and account for the effect of a light-element stellar atmosphere on
the emergent spectrum.  The strong NS surface gravity essentially
guarantees that the photosphere will be dominated by the lightest
element present (Alcock \& Illarionov 1980; Romani 1987).  For such
atmospheres with $T\sim 10^6$ K, the $E^{-3}$ dependence of free-free
absorption will shift the peak of the emission blueward from that of
an ideal BB (Romani 1987; Zampieri et al. 1995).  Previous authors have
shown that neglecting this effect and fitting the Wien tails of such
spectra with an ideal BB model significantly overestimates the
effective temperature and underestimates the emitting area (Rajagopal
\& Romani 1996; Zavlin, Pavlov, \& Shibanov 1996; Rutledge et
al. 1999).  To consider this possibility, we employed two models: 
the simple analytic spectrum emerging from the power-law atmosphere of
Heyl \& Hernquist (1998a; hereafter HH98a), with $\gamma=3$ as
appropriate for a light-element atmosphere; and the more detailed H
atmosphere model of Zavlin, Pavlov, \& Shibanov (1996; hereafter
ZPS96), developed for a weakly magnetized NS.   

For all these models, we included the effect of photoelectric
absorption by neutral gas along the line of sight (Morrison \&
McCammon 1983), which is a significant effect at low ($\lesssim 1$
keV) energies.  With the relatively small number of low-energy X-ray
photons detected from the Cas~A point source, it is difficult to
constrain $N_{\rm H}$ precisely.   Moreover, its value in 
this situation is highly spectral-model--dependent and strongly
covariant with the overall flux normalization.  However, $N_{\rm
H}$ may also be determined from radio measurements of the atomic and
molecular hydrogen column densities using the 21~cm \ion{H}{1} and 18~cm
OH absorption lines, respectively.  Keohane, Rudnick, \& 
Anderson (1996) have used such data to derive a spatially-resolved
column density map of the Cas~A SNR at 30 arcsec resolution.  Using
this map, we estimate that the column density towards the Cas~A point
source is $N_{\rm H}=(1.1\pm 0.1) \times 10^{22}$ cm$^{-2}$.  However,
this estimate must be used with caution, as the radio maps give no
information on small--angular-scale variations in the column. 
For each spectral model we employed, we fit the data both with $N_{\rm
H}$ held fixed at this value and with $N_{\rm H}$ as a free
parameter. 

Our spectral fitting results are summarized in Table 2, and a typical
model fit is shown in Figure 2.  All of the models gave formally
acceptable fits\footnote{The fact that the reduced $\chi^2$ values are
less than unity in some cases does not imply that the measurement
uncertainties (given by Poisson statistics) are underestimated.  The
probability of obtaining a reduced $\chi^2$ as small as 0.7 with only
19 degrees of freedom (assuming the true model is being fit) is 18\%,
which is thus within the 1$\sigma$ range for an acceptable fit.}; we
are unable to reject any of them on the basis of these data.  However,
the BB and NS atmosphere models fit slightly better to the data than
the other models, although with inferred radii considerably smaller
than the 10 km typically assumed for a NS radius.
\begin{deluxetable}{llcccccc}
\setlength{\tabcolsep}{0.04in}
\tabletypesize{\small}
\tablewidth{0pt}
\tablecaption{X-ray Spectral Fits for Cas~A Point Source}
\tablehead{
 &\colhead{$N_{\rm H}$}& \colhead{Photon} &  &
   \colhead{$kT^{\infty}~$\tablenotemark{c}} &
   \colhead{$R_{\rm bb}^{\infty}$~\tablenotemark{d}}& & \\
\colhead{Model}&\colhead{($10^{22}$ cm$^{2}$)} &
\colhead{index\tablenotemark{a}}&\colhead{$C_1$\tablenotemark{b}}&
   \colhead{(keV)} & \colhead{(km)} &
   \colhead{$L_{{\rm x},33}$\tablenotemark{e}} &
   \colhead{$\chi^2_{\rm red}$/dof}}
\startdata
Power law
   & $1.68^{+0.39}_{-0.22}$ & $3.13^{+0.50}_{-0.30}$ &$1.62^{+1.22}_{-0.47}$ &
       \nodata & \nodata  & $43^{+122}_{-22}$ &
       1.12/18 \\
   & 1.1 (fixed)& $2.35\pm0.12$ & $0.65^{+0.07}_{-0.05}$ &
       \nodata &\nodata & $7.4^{+1.5}_{-1.0}$ &
       1.44/19 \\
\\
Thermal brems.
   & $1.39^{+0.23}_{-0.13}$ & \nodata & $1.37^{+0.61}_{-0.40}$ &
        $1.47^{+0.34}_{-0.26}$ & \nodata & $5.2^{+3.3}_{-1.9}$ &
        0.92/18 \\
   & 1.1 (fixed) & \nodata & $0.84 \pm 0.09$ &
        $1.97^{+0.26}_{-0.21}$ & \nodata & $3.8^{+0.7}_{-0.6}$ &
        1.00/19 \\
\\
Ideal blackbody
  & $0.84 \pm 0.15$ & \nodata & \nodata &
        $0.53\pm0.04$ & $0.41^{+0.08}_{-0.07}$ & $1.7^{+1.6}_{-0.9}$ &
        0.69/18 \\
   & 1.1 (fixed) & \nodata & \nodata &
        $0.49\pm0.02$ & $0.52^{+0.05}_{-0.04}$ & $2.0^{+0.8}_{-0.6}$ &
        0.77/19 \\
\\
NS atm. (HH98a\tablenotemark{f}\ )
   & $0.85^{+0.18}_{-0.15}$ & \nodata & \nodata &
        $0.42 \pm 0.03$ & $0.67 \pm 0.12$ & $1.8^{+1.0}_{-0.6}$&
        0.70/18 \\
   & 1.1 (fixed) & \nodata & \nodata &
        $0.38 \pm 0.02 $ & $0.88 \pm 0.07$ & $2.1^{+0.5}_{-0.4}$&
        0.76/19 \\
\\
NS atm. (ZPS96\tablenotemark{g}\ )
   & $0.92^{+0.20}_{-0.16}$ & \nodata & \nodata &
        $0.28\pm0.03$ & $1.80^{+0.55}_{-0.35}$ & $2.5^{+1.6}_{-0.9}$ &
        0.74/18 \\
   & 1.1 (fixed) & \nodata & \nodata &
        $0.26\pm0.02$ & $2.23^{+0.24}_{-0.19}$ & $2.8^{+0.6}_{-0.5}$ &
        0.73/19 \\
\enddata
\tablenotetext{a}{Photon index, defined such that the unabsorbed
photon number flux $dN/dE\propto E^{-\Gamma}$.}
\tablenotetext{b}{Unabsorbed flux density at 1 keV, in units of
$10^{-3}$ photon cm$^{-2}$ s$^{-1}$ keV$^{-1}$.}
\tablenotetext{c}{For BB and NS atmosphere models, as measured by an
observer at infinity assuming a 10 km NS radius.}
\tablenotetext{d}{Implied blackbody radius assuming a source distance of 3.4
kpc, as measured by an observer at infinity.}
\tablenotetext{e}{0.1--10 keV luminosity in units of
$10^{33}$ erg s$^{-1}$ assuming a source distance of 3.4 kpc or bolometric
luminosity at infinity for blackbody models.}
\tablenotetext{f}{Analytic NS power-law atmosphere of Heyl \&
Hernquist (1998a), with $\gamma=3$.}
\tablenotetext{g}{H atmosphere for a non-magnetic NS (Zavlin et al. 1996).}
\end{deluxetable}

Interpretation of our results depends critically upon the reliability
of the energy and effective area calibrations of {\em Chandra}, which
are still being evaluated by the instrument teams.  To verify the
robustness of our conclusions, we made a rough check of these
calibrations by fitting the X-ray spectral data from the 1999 Aug 23 ACIS/S3
observation of SNR E0102$-$72 (ObsId 1231), the brightest supernova
remnant in the Small Magellanic Cloud (SMC; Hayashi et al. 1994).  For
comparison, we also analyzed archival spectral data for the same
source as observed by {\em ASCA} on 1993 May 12--13.  For both data sets,
we restricted our analysis to 0.6--2.6~keV (an energy band that both
contains the majority of the SNR flux and spans a range similar to
that of the Cas~A point source) and fit the same plasma model, with
fixed column density and non-solar abundances.  The derived plasma
temperatures agree to better than 2\%, and the overall normalizations
to within 13\%, using the preliminary calibrations. 

\subsection{X-Ray Timing}

We examined two on-axis observations of Cas~A made with the HRC as
part of the OAC program, using the HRC-S and HRC-I detectors
respectively.  These observations are summarized in Table~1.  We
obtained the processed level 2 event data from the {\em Chandra} Data
Archive.  In both cases, the overall count rate ($\sim 140$ count
s$^{-1}$) from Cas~A was below the telemetry threshold ($\approx$ 184
count s$^{-1}$) where deadtime effects become significant. For each
observation, we extracted all events within 3 arcsec of the point source
centroid to conservatively ensure that all source photons were included.
The photon arrival times, provided in terrestrial time (TT)
at the spacecraft, were corrected to barycentric dynamical time (TDB)
at the solar system barycenter using the JPL DE200 solar system
ephemeris (Standish et al. 1992) and a geocentric spacecraft
ephemeris.

We binned the events into 0.5 ms time bins and computed a Fourier power
spectrum of the resulting time series.  No significant pulsations were
detected; the highest peak in the power spectrum had a significance of
only 2$\sigma$ when the number of trials is accounted for.  To improve
our statistics, we also computed an incoherent power spectral sum of 9
ks segments of each observation.   Again, no significant pulsations
were found, with the highest peak having a significance of only 0.8$\sigma$
including the number of trials.   A large pulse frequency derivative
might spread a coherent pulsation over multiple power spectral bins,
reducing our sensitivity.   However, for the $\approx 10$ ks
observations used here, a pulse frequency derivative of $\dot\nu =
1/T^2 \gtrsim 10^{-8}$ Hz~s$^{-1}$ would be required.  This is 25
times larger than $\dot\nu$ for the Crab pulsar, and $\gtrsim 10^5$
larger than the typical value for an AXP or SGR.   To enhance our
sensitivity to non-sinusoidal pulse shapes, we also performed 
harmonic folds of the power spectrum (see, e.g., de Jager, Swanepoel,
\& Raubenheimer 1989).  No significant pulsations were detected. 

After our initial analysis was completed, we learned of a wiring problem
in the HRC on-board electronics that causes HRC events to be
time-tagged incorrectly (S.~S. Murray 2000, private communication;  
Seward 2000).  The net effect of the problem is that the timing
accuracy of the HRC is degraded from 16 $\mu$s to $\sim$4 ms.
Consequently, we conservatively conclude that the HRC analysis
presented here has no sensitivity to pulsations with frequency above
50~Hz.   We used our incoherent power spectral sum to estimate an
upper limit for the sinusoidal pulsed fraction of the Cas~A point source,
accounting for the suppression of power spectral sensitivity at higher
frequencies due to binning of the data (e.g., van der Klis 1989).  We
find that the 3$\sigma$ upper limit on the sinusoidal pulsed fraction
is $<$35\% for $\nu < 50$ Hz.  

Our sensitivity to rapid pulsations depends upon an accurate
correction for the motion of the spacecraft with respect to the
barycenter.  As a check of our barycenter corrections and of the
spacecraft ephemeris, we analyzed a 1999 August 31 HRC-I observation
(ObsID 132) of PSR B0540--69, a young 50~ms pulsar associated with the
supernova remnant N158A in the Large Magellanic Cloud, using the same
data analysis procedure as for Cas~A.  (Detailed independent analyses
of this observation are presented by Gotthelf \& Wang 2000 and Kaaret
et al. 2000.)  We compared this measurement with a 1999 September 1
observation of an overlapping field with the {\em Rossi X-Ray Timing
Explorer} (these data were generously made available by F.~E. Marshall
of NASA Goddard Space Flight Center).  The frequencies measured in the
two data sets were consistent within the uncertainties, and also
agreed with the value extrapolated from a timing model based on a 1996
{\em BeppoSAX} observation (Mineo et al. 1999).

\section{DISCUSSION}

We have shown that the X-ray point source in Cas~A has a spectrum
well-described by either an absorbed power law with photon index
2.8--3.6 and unabsorbed 0.1--10 keV luminosity (7--160)$\times 10^{33}$
erg s$^{-1}$, or an absorbed BB or modified BB with $kT^\infty\approx
0.25$--0.5 keV and bolometric luminosity $L^\infty\approx
(1$--5)$\times 10^{33}$ erg s$^{-1}$.  Our {\em Chandra} spectral and
timing measurements, combined with pre-existing limits at other
wavelengths, severely constrain plausible models for the nature of the
X-ray point source in Cas~A.   

We begin by pointing out, for completeness, that our steep power law
spectral fit essentially rules out the possibility that the point
source is a background galaxy, as active galactic nuclei (AGN)
typically lie within a range of photon indices $\Gamma=1.2$--2.2
(Turner \& Pounds 1986).  The AGN scenario is also extremely
implausible given that the point source is located within a few arcsec
of the expansion center of the SNR (Tananbaum 1999).  Moreover, given
the surface density of AGNs on the sky at this flux level (Boyle et
al. 1993), the probability of a chance coincidence is negligible.
There is thus extremely strong inferential evidence that the 
point source is indeed associated with the supernova remnant.   

We now summarize the implications of our results for several other
interpretations.

\subsection{Classical young pulsar}

If the Cas~A point source is a classical young pulsar (the
conventional product expected for a type II SN explosion), then the
X-ray radiation should be predominantly non-thermal power-law emission
from relativistic acceleration of $e^+e^-$ pairs in the corotating NS
magnetosphere (see Romani 1996).  In Table~3, the properties of the
point source are compared with the six young ($<10^4$ yr) classical
pulsars whose X-ray spectra are well measured.  All six of these pulsars
have power law X-ray spectra with $\Gamma=1.1$--1.7.  The clearest
distinction is that the spectral shape of the Cas~A point source is
considerably steeper than that of the young pulsars, although its
X-ray luminosity is marginally consistent with the lower end of the
young pulsar range.  The upper limit on the point source's X-ray pulse
fraction ($<35$\%) is lower than at least three of the measured pulsed
fractions in the young pulsars.   In the other three pulsars, the {\em
lower} limit on the pulsed fraction is consistent with the point
source's upper limit, but it is likely that these pulsar lower limits
are a drastic underestimate due to the presence of plerionic emission.
Radio pulsations have not been detected from Cas~A, with 
an upper limit of $L_{600}<530\,d_{3.4}^2$ mJy kpc$^2$ for the 600~MHz
radio luminosity (Lorimer, Lyne \& Camilo 1998), in contrast to five
of the six young pulsars\footnote{One of these pulsars, PSR
J1846--0258, was only recently discovered in the X-ray band (Gotthelf
et al. 2000).  A radio search has not been reported since the X-ray
pulsations were detected.}.  This non-detection in Cas~A could be
explained by beaming, judging from the observed radio luminosities in
young pulsars ($\approx 900$~ mJy\,kpc$^2$ for the Crab and PSR
B0540--69, but only $\approx 30$~mJy\,kpc$^2$ for PSR B1509--58, PSR
J1119--6127, and PSR J1617--5055; Taylor, Manchester, \& Lyne 1993;
Kaspi et al. 1998, 2000).  The {\em Chandra} images also show no
obvious evidence for a plerion surrounding the point source. Plerions,
powered by synchrotron emission, have been detected around five of the
six classical young pulsars in Table~3 (the exception is PSR
J1617--5055; Kaspi et al. 1998), and even around some X-ray point
sources in SNRs which are not known pulsars.  Based principally on the
X-ray spectral shape, with some support from the other properties, 
we conclude that the point source in Cas~A is not a classical young
pulsar.  
\begin{deluxetable}{llccccccccccc}
\tabletypesize{\tiny}
\setlength{\tabcolsep}{0.02in}
\tablewidth{0pt}
\tablecaption{Cas~A and Comparison Objects}
\tablehead{
 & & & & & & \multicolumn{2}{c}{Power law spectrum} &
  \multicolumn{3}{c}{Blackbody spectrum} & & 
\\ \cline{7-8}\cline{9-11}
 & & & \colhead{$P_{\rm spin}$} & 
   \colhead{$\log \tau_{\rm c}$/$\log \tau_{\rm SNR}$} & \colhead{$d$} & 
   \colhead{Photon} & \colhead{$\log L_{\rm pl}$} & \colhead{$kT^{\infty}$} & 
   \colhead{$R_{\rm bb}^{\infty}$} & \colhead{$\log L_{\rm bb}^{\infty}$} & 
   \colhead{Pulse} & 
\\
   \colhead{Source} & \colhead{SNR} & \colhead{$\beta$} & \colhead{(s)} & 
   \colhead{(yr)} & \colhead{(kpc)} & \colhead{index} & 
   \colhead{(erg/s)} & \colhead{(keV)} & \colhead{(km)} &
   \colhead{(erg/s)} & \colhead{frac.(\%)} & \colhead{Ref}
}
\startdata
\multicolumn{4}{l}{\underline{\em Cassiopeia A}} \\
Point source\tablenotemark{a} & Cas A & 0.0 & \nodata & \nodata /2.5 
    & 3.4 & 2.2--3.6 & 33.8--34.6 &
    0.5 & 0.5 & 33.3 & $<35$ & \\ 
\\
\multicolumn{4}{l}{\underline{\em Other non-plerionic X-ray point sources in 
SNRs\tablenotemark{b}}} \\
1E 0820$-$4247 & Pup A          &0.1 & 0.075?  & 3.9/3.6 & 2.0  & \nodata & 
    \nodata & 0.28 & 2  & 33.6  & 20? & 1--3 \\
1E 1614$-$5055  & RCW 103        &0.0 &\nodata  &  \nodata/3.1 & 3.1 &
    \nodata & \nodata & 0.56 & 0.81 & 33.9 & \nodata & 4,5 \\
1E 1207$-$5209 & PKS 1209$-$52  &0.2 & 0.424  & \nodata /3.8 & 1.5 &
    \nodata & \nodata &0.25 & 1.1 & 33.1  & \nodata & 6--8 \\
\\
\multicolumn{4}{l}{\underline{\em Young classical pulsars}} \\
PSR J1846$-$0258 & Kes 75 & $<0.3$ & 0.325 & 2.9/3.0 & 19 & 1.1 & $>34.6$ & 
    \nodata & \nodata & \nodata & $>6$ & 9 \\
PSR B0535$+$21  & Crab &0.0 & 0.033 & 3.1/3.0 & 2.0 & 1.74 & 36.2 &
    \nodata & \nodata & \nodata & $\gtrsim 75$ & 10 \\
PSR B1509$-$58  &MSH 15$-$5{\it 2} &0.2 & 0.150 & 3.2/3.2 & 5.2 & 1.36 &
    35.3 & \nodata & \nodata & \nodata & 65 & 11,12\\
PSR B0540$-$69 & N158A&0.2 & 0.050 & 3.2/3.1 & 47 & 1.3 & 36.6 &
    \nodata & \nodata & \nodata & $\gtrsim 15$ & 13,14 \\
PSR J0537$-$6910 &N157B&0.5 &0.016 & 3.7/3.7 & 47 & 1.6 & 35.5 &
    \nodata & \nodata & \nodata & $>10$ & 15--17 \\
PSR J1617--5055 &\nodata& \nodata & 0.069 & 3.9/\nodata & 6.5 & 1.6& $>34.5$ & 
    \nodata & \nodata & \nodata & $\sim$50 & 18,19
\\
\multicolumn{4}{l}{\underline{\em Anomalous X-ray pulsars}} \\
1E 1841$-$045  & Kes 73 &0.2 & 11.8 & 3.6/3.3  & 7.0 & 3.4 & 36.9 & 
     \nodata & \nodata & \nodata & 15\tablenotemark{c} & 20,21 \\
1E 1048.1$-$5937  & \nodata &\nodata & 6.45 & 3.7?/\nodata  & 3.0 & 2.5 &
     34.4 & 0.64 & 0.6 & 33.9 & 70 & 22,23 \\
AX J1845$-$0258\tablenotemark{a} & G29.6+0.1 &0.2 & 6.97 &?/$<$3.9 & 8.5 &
    4.6 & 38.6 & 0.72 & 1.6 & 34.9 & 50 & 24,25 \\
RXS J1708$-$40    & \nodata &\nodata & 11.0 & 4.0/\nodata  & 10.0 & 2.9 &
     36.9 & 0.41 & 8.9 & 35.5 & 50 & 26--28 \\
4U 0142$+$61      & \nodata &\nodata & 8.69 & 4.8?/\nodata & 1 & 3.7 & 35.9 & 
     0.39 & 2.4 & 33.5 & 10 & 29,30 \\
1E 2259$+$586  & CTB 109 &0.3 & 6.98 & 5.3/4.0  & 4.0 & 4.0 & 36.9 &
     0.43 & 2.2 & 34.3 & 30 & 28,31 \\
\\
\multicolumn{4}{l}{\underline{\em Soft gamma repeaters\tablenotemark{d}}} \\
SGR 1900$+$14  & G42.8+0.6? &1.2 & 5.16 & 2.8?/4.0  & 5.0 & 1.1  & 34.5 & 
     0.51 & 1.4 & 34.2 &  11 & 32,33 \\
SGR 1806$-$20  & \nodata\tablenotemark{e} &0.0 & 7.47 & 3.1?/\nodata & 
     14.5 & 2.2 & 36.0 & \nodata & \nodata & \nodata &  23 & 34--36  \\
SGR 0526$-$66  & N49 &0.6 & 8 & ?/3.7 & 47 & \nodata & \nodata & 
     \nodata & \nodata & \nodata &  $<$66&  37,38 \\
SGR 1627$-$41  & G337.0$-$0.1? &2 & 6.41? & ?/3.7 & 11.0 & 2.5  & 35.8  & 
     \nodata & \nodata & \nodata &  10  & 39--41  \\
\enddata
\tablenotetext{a}{For the Cas~A point source and AX~J1845$-$0258, 
the powerlaw (PL) and (blackbody) BB models are
{\em alternative} fits to the same data.  For all the other sources,
the PL and BB models are each components of a {\em combined} model.}
\tablenotetext{b}{Here, we only include those sources that are
well-established and have constrained spectral properties.}
\tablenotetext{c}{We note that this value is {\em half} that reported
in the literature, due to a different definition of pulsed fraction
by these authors.}
\tablenotetext{d}{The spectral properties and pulse fractions 
listed for the SGRs are for {\em quiescent} emission.}
\tablenotetext{e}{The extended radio nebula G10.0$-$0.3,
once identified as a SNR,
is now thought to be powered by a massive companion or the SGR itself
(Gaensler 2000; Eikenberry \& Matthews 2000; Frail, Vasisht \&
Kulkarni 1997).}
\tablerefs{
(1) Petre et al. 1982; (2) Petre, Becker \& Winkler 1996; 
    (3) Pavlov, Zavlin \& Tr\"{u}mper 1999;
(4) Helfand \& Becker 1984; (5) Mereghetti, Bignami \& Caraveo 1996; 
(6) Tuohy \& Garmire 1980; (7) Gotthelf, Petre \& Hwang 1997; 
(8) Zavlin et al. 2000;
(9) Gotthelf et al. 2000; (10) Harnden \& Seward 1984; 
(11) Marsden et al. 1997; (12) Gaensler et al. 1999a;
(13) Finley et al. 1993; (14) Seward \& Harnden 1994;
(15) Wang \& Gotthelf 1998a;  (16) Wang \& Gotthelf 1998b;
  (17) Marshall et al. 1998;
(18) Torii et al. 1998b; (19) Kaspi et al. 1998;
(20) Gotthelf \& Vasisht 1997; (21) Gotthelf, Vasisht \& Dotani 1999b;
(22) Corbet \& Mihara 1997; (23) Oosterbroek et al. 1998;
(24) Torii et al. 1998a; (25) Gaensler et al. 1999b;
(26) Sugizaki et al. 1997; (27) Israel et al. 1999a;
    (28) Kaspi, Chakrabarty \& Steinberger 1999;
(29) White et al. 1996; (30) Israel et al. 1999b;
(31) Rho \& Petre 1997; 
(32) Vasisht et al. 1994; (33) Woods et al. 1999a 
(34) Sonobe et al. 1994; (35) Corbel et al. 1997; 
   (36) Kouveliotou et al. 1998 ; 
(37) Vancura et al. 1992;  (38) Marsden et al. 1996;
(39) Woods et al. 1999b; (40) Corbel et al. 1999; 
(41) Hurley et al. 2000.
}
\tablecomments{$\beta$ is defined as the ratio between the source distance
from the center of the SNR to the radius of the SNR; 
the characteristic age is defined as $\tau_{\rm c} = P/2\dot{P}$; 
luminosities are either bolometric (BB) or for the 
$0.1-10$~keV band (PL).}
\end{deluxetable}

If the Cas~A point source is a NS, the absence of both detectable radio
pulsations and a synchrotron nebula may indicate that it
lies beyond the so-called pulsar ``death line'', an empirical
boundary on the spin-period--magnetic-field plane beyond which radio
pulsars have generally not been detected, presumably because the NS does
not generate enough $e^+e^-$ pairs to power significant non-thermal
emission\footnote{The recent identification of a radio pulsar well
beyond the death line, PSR J2144--3933 ($P=8.51$ s, $B\approx 6\times
10^{11}$ G) indicates that this argument must be used cautiously (Young,
Manchester, \& Johnston 1999).} (Chen \& Ruderman 1993).   In this
case, we would expect a strongly magnetized ($B\gtrsim 10^{11}$ G) NS to
have a spin period of order at least a few seconds.  Conversely, a rapidly
spinning NS ($P\sim 30$ ms) would have a very weak magnetic field
($B\lesssim 10^8$ G), perhaps consistent with delayed field
growth (Blandford, Applegate \& Hernquist 1983).  

\subsection{Cooling neutron star}

An alternative interpretation is that the X-rays from the Cas~A point
source arise from thermal emission from a cooling NS.  A 300~yr old NS
cools primarily via neutrino emission; standard cooling models predict
thermal photon emission from the surface with $kT^{\infty}=0.15$--0.25
keV (Page \& Sarmiento 1996; Page 1998).  All of our BB and modified
BB fits for the Cas~A point source yield somewhat higher temperatures
($kT^\infty\approx 0.25$--0.5 keV), as well as much smaller BB radii
($R^\infty_{\rm bb}\approx0.6$--2.6 km) than expected for a 10 km NS,
even when accounting for a light-element atmosphere.  The NS
atmosphere models (HH98a, ZPS96) that we fit to the data were computed
assuming a weak ($B\lesssim 10^{10}$ G) magnetic field, which may be a
poor assumption for a young NS.  Qualitatively, however, the presence
of a magnetic field of order $\sim 10^{12}$ G will shift the peak of the
emission in a light-element atmosphere {\em redward} towards the ideal
BB case (Pavlov et al. 1995), thus exacerbating the discrepancy
between the data and standard NS cooling curves.  We note, however,
that the behavior of spectral shifts in ultrastrong magnetic fields
($B\sim 10^{14}$--$10^{15}$ G) has not yet been calculated.

While our inferred temperature is marginally consistent with standard
NS cooling curves, the small emitting area remains problematic,
especially given our limits on the X-ray pulsed fraction.
Strong ($B \sim 10^{12}$--$10^{13}$ G) magnetic fields will produce a
non-uniform temperature distribution on the surface of a NS,
owing to anisotropic electron conduction through the star's outer
envelope (e.g. Greenstein \& Hartke 1983; Heyl \& Hernquist 1998b,
2000).  However, the resulting temperature variation across the
surface does not produce small hot spots, but is instead smoothly
varying with a local flux roughly $\propto \cos ^2 \psi$, where $\psi$
is the polar angle from the magnetic axis.  This would reduce the
effective area of the emitting surface by a factor $\sim 3$, far less
than the factor of 30--100 required by our spectral fits.

Similar conclusions motivated Pavlov et al. (2000) to consider a model
in which the magnetic polar caps are intrinsically hotter than the
bulk of the stellar surface, as a result of horizontal chemical
abundance gradients through the conductive envelope.  Light element
envelopes transmit heat more readily than ones made of heavy elements
(Chabrier, Potekhin \& Yakovlev 1997; Heyl \& Hernquist 1997a), so hot
polar caps consisting of hydrogen embedded in a cooler iron crust can,
in principle, yield an emitting area consistent with the spectral fits
(Pavlov et al. 2000).  However, such a model predicts that the
emission should be pulsed at the rotation period of the star, unless
either the magnetic and rotation axes are nearly aligned, or the line
of sight nearly coincides with the rotation axis.  Pulsed fractions at
the level of 10--70\% required to account for the putative thermal
emission from middle aged radio pulsars (e.g. Becker \& Trumper 1997)
and AXPs (e.g., Mereghetti 2000) can be produced from {\it smoothly}
varying properties of NS envelopes, such as anisotropic heat
conduction (Heyl \& Hernquist 1998b) or directionally dependent
opacities (e.g. Pavlov et al. 1994, Zavlin et al. 1995, Shibanov et
al. 1995), even when gravitational bending of light is included
(e.g. Page 1995, Heyl \& Hernquist 1998b).  An even larger pulsed
fraction will result, in general, from the hot spot model.  For
example, in the case of the orthogonal rotator model of Pavlov et
al. (2000), we estimate typical maximum to minimum flux variations $>
2$ for $1.4 M_\odot$ NSs with radii $R>7$ km.  This is in
severe disagreement with the upper limits we obtain for the pulsed
fraction.\footnote{For special choices of the NS radius, however,
gravitational bending of light will make the entire stellar surface
singly visible, eliminating any pulsed component; see, e.g. figure 7
of Heyl \& Hernquist (1998b).}  Observationally, we cannot yet exclude
the case of a nearly aligned rotator, but it is not clear that the
horizontal abundance gradients required by the Pavlov et al. model
will be stable for long times in the liquid crust.

\subsection{Accretion onto a neutron star or black hole}

We now consider the possibility that the point source in Cas~A is
powered by accretion onto a NS or a BH.  This possibility was
also raised by Umeda et al. (2000) and Pavlov et al. (2000).  We assume
that the accretion is fed by fallback material left over after the
original supernova explosion (e.g. Chevalier 1989).  We prefer such a
model to a binary accretion model since there is no optical/IR
evidence for a such binary companion star (van den Bergh \& Pritchet
1986).  A very low-mass dwarf companion might have evaded detection,
but such a companion would be unlikely to remain bound in the binary
following the supernova explosion, given the high mass of the Cas~A
progenitor (Brandt \& Podsiadlowski 1995; Kalogera 1996). 

We begin with the possibility of accretion onto a NS.  If the
accretion occurs via a thin disk extending down to the marginally
stable orbit at $6GM/c^2$, or the NS surface (whichever is 
larger), then we expect significant emission from an equatorial
boundary layer where the accreting material meets the star.  The
emitting zone is expected to have a radial extent roughly equal to the
local scale height of the disk (Narayan \& Popham 1993; see also
Inogamov \& Sunyaev 1999 for a recent discussion of boundary layer
models).  If the boundary layer is optically thick, then it will emit
BB radiation.  For an accretion luminosity $\sim10^{33.5} ~{\rm
erg\,s^{-1}}\lesssim10^{-4}L_{\rm Edd}$, the scale height is $\sim0.1$ km,
and the effective area of the radiating zone is $\sim$few km$^2$.
This is in reasonable agreement with the effective area $\sim$km$^2$
determined from fitting the Chandra data (Table 2).  We consider this
a viable model, although we note that the optical thickness of such
boundary layers is not well understood.  

This boundary layer model requires that the NS have a very weak
magnetic field.  Specifically, the magnetospheric radius $r_{\rm m}$
has to be smaller than the stellar radius, which implies that $B<10^7$
G for the estimated X-ray luminosity.  If the magnetic field is
somewhat stronger, then the disk would be terminated at $r_{\rm m}$. 
Even in this case, we might expect a boundary layer to develop at
$r_{\rm m}$, and the model would be consistent with the observations
for fields $B\lesssim 10^9$ G.   We note that for sufficiently large $B$,
accreting material is centrifugally expelled from the system at
$r_{\rm m}$ (Illarionov \& Sunyaev 1975).  In this case, very little
material would reach the neutron star and there would be negligible
X-ray emission (e.g., Chatterjee, Hernquist, \& Narayan 2000; Alpar
1999, 2000).  

If the NS instead has a magnetic field $\gtrsim10^{12}$ G typical of
young radio pulsars, then boundary layer emission is unlikely to
explain the observed X-rays.  In this case, we would have to assume
that the accreting material is able to reach the surface of the
neutron star, which would require that the neutron star be spinning
quite slowly (in order to avoid centrifugal expulsion of matter).  The
effective area of the radiating zone would again be small, and could
thus be consistent with the observations.  However, we would predict
strong coherent pulsations in the X-ray signal .  Our upper limit of
25\% on the pulsed fraction from the Cas~A point source does not rule
out pulsations at the level observed in many known $B\sim 10^{12}$ G
accreting pulsars.  More sensitive searches for pulsation would be
very worthwhile.

A third kind of accreting NS model is one in which the accretion disk
is truncated at a large transition radius, and the flow farther in
occurs via an ADAF (as in the Narayan et al.  1997 model for
low-luminosity BH binaries).  Boundary layer emission in such a model
is not well understood, so we are not in a position to predict the
emission spectrum.  However, if the ADAF is terminated at $r_{\rm m}$
and the material then flows onto the magnetic poles, the spectrum
would be similar to the case with a thin disk which was discussed above.

Accreting BH models face considerably greater difficulty in fitting
the observations.  This is because neither a boundary layer nor
channeled accretion onto magnetic poles is expected.  Therefore, any
blackbody emission should be primarily from the inner accretion disk,
with an effective area of several times $\pi R_{\rm S}^2$, where
$R_{\rm S}$ is the Schwarzschild radius of the black hole.  For a
$10M_\odot$ BH, the area is $\sim10^3$ km$^2$, which is clearly
inconsistent with the X-ray data.  This means that the X-ray emission
from a BH point source in Cas~A would have to originate either via
Compton scattering in an optically thin corona over a thin disk, or
via optically thin bremsstrahlung emission from a hot ADAF.  We have
explored models of this kind, using the modeling techniques described
in Narayan et al. (1997) and Quataert \& Narayan (1999), but we find
that we need to fine-tune the models to an uncomfortable degree to fit
the observations.

It is important to note that accretion models are severely constrained
by the observed flux ratio between X-ray and optical bands.  The
optical limits on a stellar remnant in Cas ~A are $I\gtrsim23.5$ and
$R\gtrsim24.8$  (van den Bergh \& Pritchet 1986).  Applying the
extinction corrections estimated by these authors, the
X-ray--to--optical flux ratio is $F_{\rm X}/F_{\rm opt}>100$.
Although this is typical for bright X-ray binaries, it is considerably
larger than the ratios observed in quiescent accreting BHs and NSs
with luminosities comparable to the Cas~A point source, e.g. $F_{\rm
X}/F_{\rm opt}\sim 1/30$ for the BH system A0620--00 and $F_{\rm
X}/F_{\rm opt}\sim 1/3$ for the NS system Cen X-4 (McClintock \&
Remillard 2000).   Both sources have an X-ray luminosity lower than
the point source in Cas~A, and both have optical luminosities
significantly higher than the upper limit for the Cas~A source. If the
spectra of these sources are characteristic of accreting BHs and NSs
at low luminosities, then accretion models for Cas~A are ruled out
with high confidence.     

It is possible, however, to evade this conclusion by arguing that
these low-mass X-ray binaries (LMXBs) are a poor comparison to the
Cas~A point source.  First of all, the quiescent optical emission in
some BH binaries has been modeled as synchrotron emission from a hot
advection-dominated accretion flow (ADAF) close to the BH (Narayan,
McClintock \& Yi 1996; Narayan, Barret \& McClintock 1997).  However,
if the emission instead comes from a ``hot spot'' (e.g., where the
accretion stream from the mass donor hits the thin accretion disk),
then it is specific to binary systems and should be absent in systems
with a fallback disk.  This possibility has not been modeled in any
detail.  Also, X-ray reprocessing in the accretion disk is a
significant source of optical emission in LMXBs (van Paradijs \&
McClintock 1995), especially since the outer disks evidently subtend a
large semi-angle ($\sim 12^\circ$) at the central object (de Jong, van
Paradijs, \& Augusteijn 1996).  The reason for such large angles may be
warping of the outer disk (see Dubus et al. 1999; Esin et al. 2000).
But if such warps are induced primarily by binary effects, then a
fallback disk in Cas~A might not be similarly warped (although it has
been argued that irradiation alone can lead to significant warping;
see Petterson 1977; Pringle 1996; Maloney et al. 1996).  For an
unwarped disk, the subtended semi-angle would be set by the disk's
scale height at the relevant radius ($\sim 10^{10}$ cm).  By employing
an analysis similar to that in Perna, Hernquist \& Narayan (1999), we
estimate that the re-emitted optical flux in such a model would be
(just) below the optical flux limits for the Cas~A source.

Thus, the optical limits strongly constrain accretion models, but do
not yet conclusively eliminate them.  By comparing the predictions of the
models with the X-ray observations, specifically the small effective
area $\sim {\rm km^2}$ of the emission, we conclude that models with
accreting NSs are more promising than those with BHs.  

\subsection{Comparison with AXPs, SGRs, and radio-quiet point sources}

It is interesting to compare the properties of the Cas~A point source
with three classes of objects whose nature remains puzzling: AXPs,
SGRs, and the radio-quiet non-plerionic X-ray point sources in SNRs.
Both AXPs and SGRs pulse with slow spin periods despite being young
objects (based on their association with SNRs).  We summarize the
X-ray properties of these objects in Table~3.  

The X-ray spectrum of most AXPs is best characterized by a
two-component spectrum consisting of a $\sim 0.5$ keV BB and a steep
($\Gamma=3$--4) power law extending out to 10 keV, with comparable
luminosity in both components.  The BB components have $R^\infty_{\rm
bb}=1$--4 km and $L^\infty_{\rm bb}\sim 10^{34}$--$10^{35}$ erg
s$^{-1}$, quite similar to the BB fit for the Cas~A point source.  (We
note that, due to poor photon statistics in the present data, we would
be unable to detect the presence of an additional power-law component
with this BB fit if the Cas~A point source has an AXP-like
two-component spectrum.)  While four of the AXPs have large pulsed
fractions (30--70\%), two have much lower pulsed fractions (10--15\%)
which are consistent with the non-detection of pulsations in
Cas~A. However, the X-ray luminosity of the Cas~A source is at least a
factor of three to ten lower than that of the AXPs, in spite of its
relative youth.  Overall, the properties of the Cas~A point source are
roughly consistent with being an underluminous AXP.  This possibility
can be tested by a deeper search for long-period X-ray pulsations.

The quiescent emission of SGRs may be fit with a power law with
$\Gamma=1.1$--2.5 and a 2--10 keV luminosity of (3--100)$\times
10^{34}$ erg s$^{-1}$.  These spectral parameters are also consistent
with those measured for Cas~A.  However, we note that the proposed
associations between SGRs and SNRs are somewhat tenuous, in contrast
to those proposed for AXPs (Gaensler 2000). Moreover, no soft
gamma-ray bursts have ever been detected from the direction of Cas~A.

Finally, there are the three radio-quiet X-ray point sources that each
lie near the center of an SNR and have no evidence for a plerion.
The X-ray properties of these sources are summarized in Table~3.  The
X-ray emission from these objects is well characterized by a BB spectrum
with $kT^\infty\approx 0.2$--0.6 keV and $R_{\rm bb}\approx0.6$--3
km, comparable to what we measure in Cas~A.  From a spectral point of
view, these four sources form a remarkably homogeneous group.
Pulsations ($P=424$ ms) have recently been detected from the source in
PKS 1209--52 (Zavlin et al. 2000), and possibly from the source in
Pup~A (Pavlov, Zavlin, \& Tr\"umper 1999); in both cases, the measured
pulsed fraction is below the limit placed in Cas~A.  However, the point  
source in RCW 103 has shown order of magnitude variability in its flux 
(Gotthelf, Petre, \& Vasisht 1999a), with a possible 6~hr periodicity
recently reported (Garmire et al. 2000).  This is in stark contrast to
the Cas~A point source, which shows no evidence for strong variability
on either short ($\sim$hours--weeks; see Table 1) or long ($\sim$years;
see Pavlov \& Zavlin 1999; Pavlov et al. 2000) time scales.  

\section{CONCLUSIONS}

At present, we do not have a unique model to account for the observed
properties of the X-ray point source in Cas~A.  Thermal emission from
an isolated, cooling NS can account for the data, but only if
unconventional modifications are incorporated into spectral models in
order to satisfy the requirement that the emitting area be small.
Localized hot spots can, in principle, be produced on the 
surface of a NS by horizontal chemical abundance gradients in the
liquid envelope (Pavlov et al. 2000).  
However, such hot spots should give rise to a significant pulsed
component to the emission, in general, and such models will likewise
be constrained by future limits or detections of X-ray modulation.
Accretion onto a BH does not appear promising, but NS accretion
provides a viable mechanism for explaining the characteristics of the
X-ray emission. If the NS is very weakly magnetized, the observed
emission could arise from a boundary layer.  Otherwise, the X-rays
could be produced by magnetically-channeled accretion onto the NS.  In
the latter case, we expect pulsed emission, so again, future timing
limits test this interpretation.  

The point source in Cas~A is similar in many respects to the AXPs and
quiescent SGRs, which have been interpreted as ultramagnetized neutron
stars (``magnetars''; Duncan \& Thompson 1992; Thompson \& Duncan
1996; Heyl \& Hernquist 1997a, 1997b) or as neutron stars with normal
magnetic fields of order $10^{12}$--$10^{13}$ G accreting from
fallback disks (Chatterjee et al. 2000; Alpar 1999, 2000).  It is important
to note the detailed thermal emission from neutron stars with magnetic
fields $B\sim 10^{14}$--$10^{15}$ has yet to be calculated.
Conceivably, X-ray spectra in this regime could account for the small
emitting areas of the Cas~A point source and the AXPs without
requiring small polar hot spots.

Regardless of the eventual resolution to the puzzling aspects of the
point source in Cas~A, it is already clear that the mere existence of
this and similar objects demands a dramatic revision of our generally 
accepted notions of the nature of compact objects found in supernova
remnants. The discovery of radio pulsars in the 1960s led to a
paradigm in which supernovae generally leave
behind strongly magnetized, rapidly rotating neutron stars.  The
subsequent failure to locate radio pulsars in the majority of SNRs has
been a long-standing problem for this point of view (Kaspi 2000,
Gotthelf \& Vasisht 2000).  The unanticipated properties of the Cas~A 
point source and the relative youth of the Cas~A SNR imply that the
birthrate of ``unusual'' compact objects is likely to be at least
roughly comparable to that of radio pulsars, potentially resolving the
difficulties posed by SNR/pulsar associations.  In this sense, the
identification of the point source in Cas~A may be as
significant to our understanding of neutron stars as was the original
discovery of radio pulsars. 

\acknowledgements{We are grateful to Fred Baganoff, Mark Bautz, Bryan
Gaensler, Jack Hughes, Vicky Kalogera, Vicky Kaspi, Herman Marshall,
Rosalba Perna, Dimitrios Psaltis, and Norbert Schulz for advice,
suggestions, and useful discussions.  We also thank Frank Marshall for
allowing us to use his proprietary {\em RXTE} data to verify our pulse
timing of PSR B0540--69, as well as Fred Seward and Arnold Rots for
help with retrieving data from the {\em Chandra} Data Archive.  The
{\em ASCA} data for SNR E0102--72 was obtained from the High Energy
Astrophysics Science Archive Research Center (HEASARC) at NASA Goddard
Space Flight Center.}

\end{document}